\documentclass[prl,twocolumn,aps,nofootinbib]{revtex4-1}
\usepackage{graphicx}
\begin{document}
\title{Approaching universality in weakly-bound three-body systems}
\author{Vladimir Roudnev and Michael Cavagnero}
\affiliation{Department of Physics and Astronomy, University of Kentucky,
Lexington, KY 40506-0055}
\date{\today}
\begin{abstract}
Atom-dimer scattering below the three-body break-up threshold is studied for a system of three
identical bosons. The atom-dimer scattering length and the energy of the most weakly-bound three-body
state are shown to be strongly correlated. An appropriate rescaling of the observables reveals
the subtlety of the correlation, and serves to identify universal trends in the unitary limit of divergent
two-body scattering length. The correlation provides a new quantitative measure of the degree
of universality in three-body systems with short-ranged interactions, as well as a consistency check
of  effective field theories and other theoretical models.
\par
\noindent
\bigskip
PACS numbers:
\end{abstract}
\maketitle
Among the most striking results of recent experimentation
with supercooled atomic gases is the demonstration that
trap loss rates are extraordinarily sensitive to
few-body interactions within a trapped many-body system.
The experiments of many groups \cite{BEC} show signatures
of few-body correlations within a trapped
ensemble of Bose alkali vapors at nano-Kelvin temperatures.
This discovery has stimulated a large number of theoretical
and experimental investigations, particularly in the
’unitarity limit’ of a divergent two-body scattering length,
where the system has no natural length scale beyond that
of the trapping potential.

It is widely anticipated that in the vicinity of unitarity,
supercooled Bose gases display universal collective
properties. However, criteria for the onset of universality
and measures of the degree of universality have not
yet emerged. The primary purpose of this Letter is to
provide such a measure through a thorough investigation
of atom-dimer scattering for a wide variety of two-body
short-ranged potentials and the zero-range interaction model.

\begin{figure}
\includegraphics[clip=true,width=0.35\textwidth]{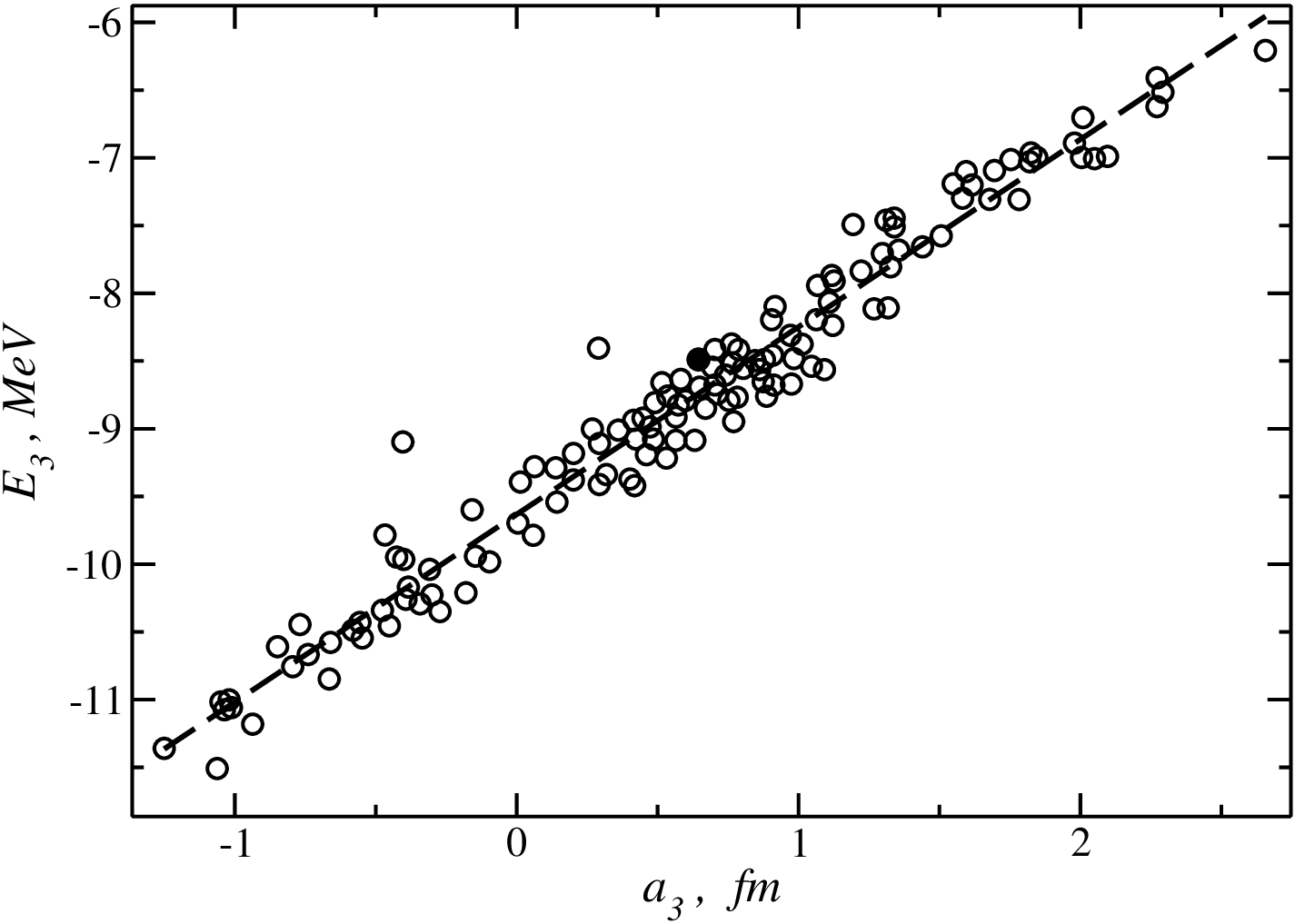}
\caption{
The Phillips line, showing the unexpected linear correlation between the triton bound state energy
and the neutron-deuteron scattering length (data from ref.~\cite{EfimovTkachenko}). }
\label{fig1}
\end{figure}

Weakly-bound few-body systems with relatively large
two-body scattering lengths have long been studied in
nuclear physics. A salient example is a result published
by Phillips in 1968 \cite{Phillips} which puzzled nuclear theorists for
more than a decade. Phillips compared results of calculations
of the neutron-deuteron scattering length and the
energy of the triton bound state made with different two-
body potentials. He found that, unlike two-body scattering
in which the energy of the last bound state scales
as the inverse squared scattering length ($E\sim-1/a^2$),
the bound state energy of the triton (n-n-p) appears
approximately proportional to the scattering length of the
neutron-deuteron collision.
The data illustrating this correlation is shown in Figure~\ref{fig1}.

Many subsequent works \cite{subsequent} confirmed that a strong
linear correlation between these two observables exists.
Two decades later a simple and elegant explanation of
the Phillips line was given by Efimov and Tkachenko
\cite{EfimovTkachenko}, who simply noted that the apparent linearity derives
from the fact that the two-body potentials used sample
only a small portion of the space of scattering parameters;
that is, they yield similar values for the neutron-deuteron
scattering lengths.

While the origin of Phillips' observation is now well understood,
modern experiments with ultracold gases are
able to sample a much wider range of scattering parameters
by magnetic field tuning through a Fano-Feshbach
resonance. This suggests that the correlation between
two- and three-body parameters can be investigated in
far greater detail than in earlier nuclear physics studies.
Theoretical and experimental works related to universality
in ultracold Bose gases have been concentrated
on three-body recombination in the close vicinity of the
three-body threshold. In this work we discuss the properties
of the three-body system at the two-body threshold
and below.

In this work we introduce a new parametrization of the relationship between the
three-body (atom-dimer) scattering length and the energy
of the last three-body bound state for the specific
case of three identical Bosons. This relationship is referred
to below as the “modified Phillips line.” In contrast
to the original Phillips line, this relationship is found to
be linear over a large range of interaction parameters
and more directly reflects the well known threshold law
for single-channel scattering, namely $E\approx\frac{1}{m a^2}$.
It also provides a simple test for universality of three-body systems.

\begin{figure}
  \includegraphics[clip=true,width=0.45\textwidth]{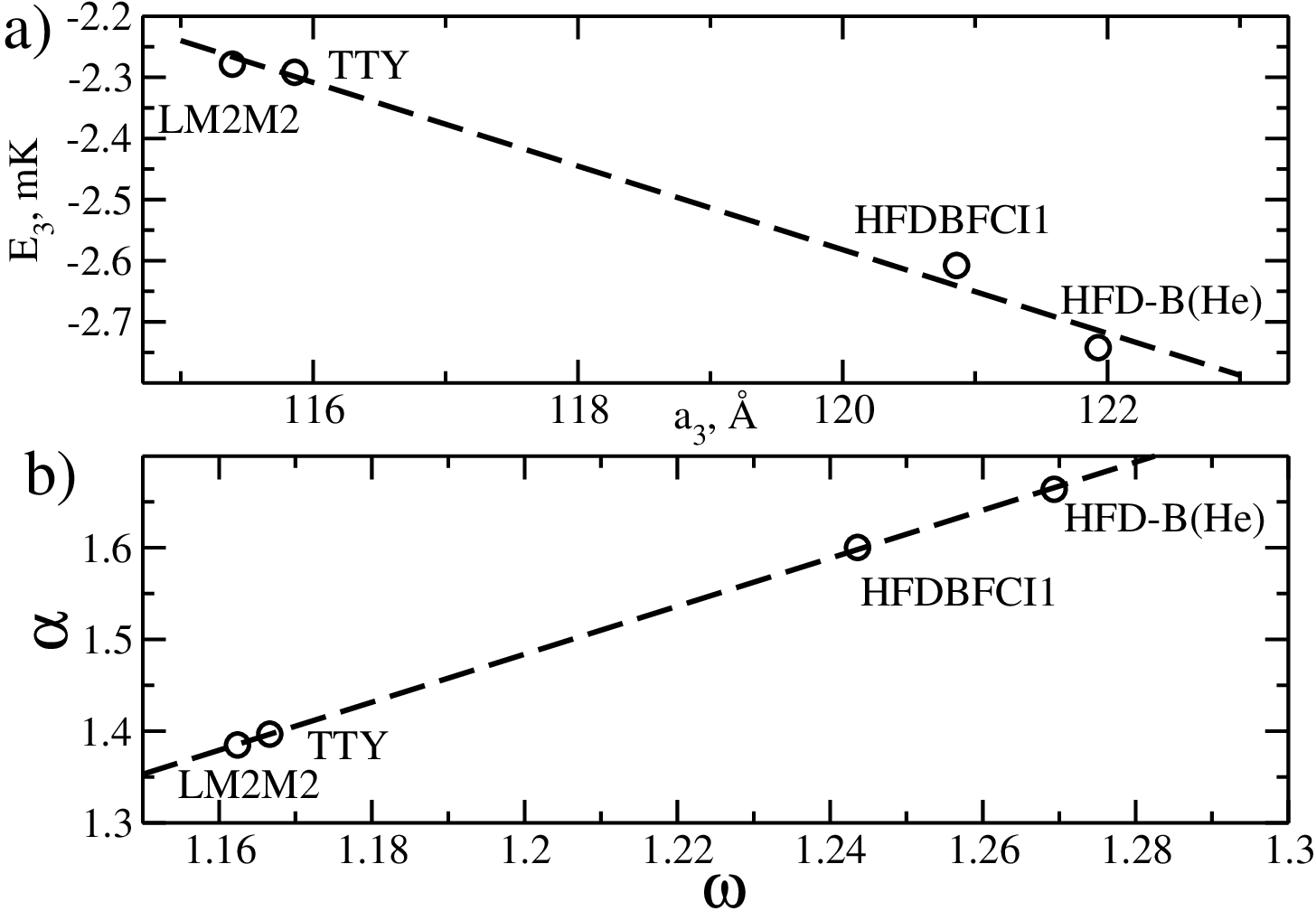}
  \caption{
       a) Phillips line for three $^4$He atoms:
       the $He_3$ binding energy as a function of the $He-He_2$ scattering length
       based on four commonly used two-body potentials. Note the approximately linear relationship (dashed line).
       b) Modified Phillips line for three $^4$He atoms for the same set of potentials. Note an improved linearity.
           }
  \label{fig2}
\end{figure}

As pointed out by Efimov and Tkachenko \cite{EfimovTkachenko},
for a weakly-bound three-body state the following approximate correlation
should hold
\begin{equation}
  E_2 - E_3 \approx 1/(2m_{12} a_3^2)
  \label{eq:prop}
\end{equation}
where $E_2$ is the dimer binding energy and $m_{12}$ is the reduced mass of
the particle-dimer system. A more transparent representation of the strength
of the correlation is obtained by rewriting Eq. \ref{eq:prop} in terms of
the scaled dimensionless variables
\begin{equation}
  \begin{array}{l}
    \alpha \equiv a_3\sqrt{-2 m_{12} E_2} \\
    \omega\equiv  1/\sqrt{E_3/E_2 -1}
  \end{array}
  \label{eq:AlphaOmega}
\end{equation}
The variable $\alpha$ can be thought of as a dimensionless scattering length,
and $\omega$ characterizes the three-body binding energy. If the three-body state
nearest to the threshold is deeply bound, $\omega$ is small; large values
of $\omega$ indicate the existence of a weakly-bound three-body state.

Using a recently developed three-body code, we have strenuously tested the
well studied case of three bosonic $^4$He atoms
%\cite{Glockle,KolganovaMotovilov,Roudnev,Roudnev1,Roudnev2,ELG,NFJ,BG,VK,Yarevsky}
\cite{He3Calc}, and so
we will use the helium trimer as a first illustration of our improved parameterization.
The traditional Phillips line for helium trimer states calculated with four commonly used
two-body potentials is plotted in Figure~\ref{fig2}a, where $E_3$ is the energy of
the $He_3$ threshold bound state (relative to the three-body break-up threshold),
and $a_3$ is the scattering length for a helium atom and a bound helium dimer.
Figure \ref{fig2}b shows the modified Phillips line obtained using the
suggested scaling with the dimer binding energy. Note that the scaling
results in an improved fit.

Equation \ref{eq:prop} derives from the fact that atom-dimer scattering is dominated
by the pole of the $t$-matrix corresponding to the near-threshold state
of the trimer. Generally speaking, we should expect the linear relation
of the rescaled parameters $\alpha$ and $\omega$ to hold providing that
no other poles of the $t$-matrix are relevant in the energy range of interest.
We have tested this assertion for systems with three indistinguishable bosons
by studying the correlation for a variety of two-body potentials with widely
adjusted two-body scattering lengths and binding energies.

\begin{figure}
  \includegraphics[clip=true,width=0.35\textwidth]{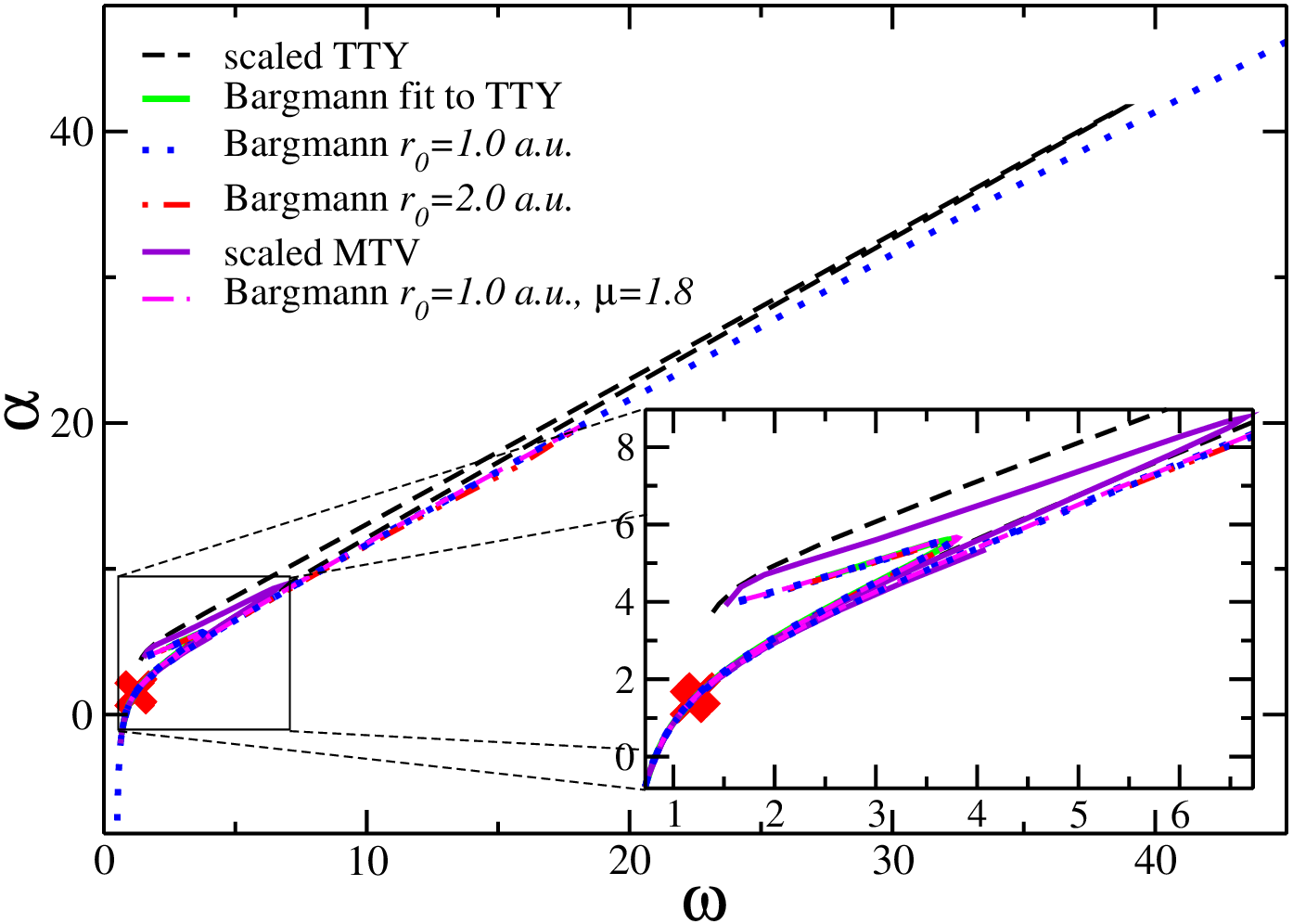}
  \caption{(Color online)
   A modified Phillips line plotted with rescaled parameters for a wide variety of two-body parameters and
   several alternative two-body potentials. Red crosses mark the results for realistic He-He potentials
   (see Fig.~\ref{fig2})}
  \label{fig3}
\end{figure}

Figure~\ref{fig3} illustrates the correlation for six different families of two-body
potentials. The potentials include the TTY potential with an artificial coupling constant,
the family of Bargmann potentials \cite{BargmannPot} with fixed small
effective range and a scattering length varying from 2 to 250 atomic units,
the Bargmann potential with an effective range simulating the He-He interaction,
the family of Bargmann potentials with varying asymptotic normalizing constant, and
the MTV potential ("symmetric model" for the triton) with a varying coupling constant.
The near-linearity over a wide range of $\omega$ is apparent in the figure, though deviations
can be seen, especially for small values of the $3-$body scattering length,
magnified in the inset of Figure~\ref{fig3}, where the complexity of the correlation is revealed.
An interesting aspect of the plot, which we had not anticipated, is the
emergence of a universal behavior as the $3-$body scattering length approaches zero.

\begin{figure}
\includegraphics[clip=true,width=0.35\textwidth]{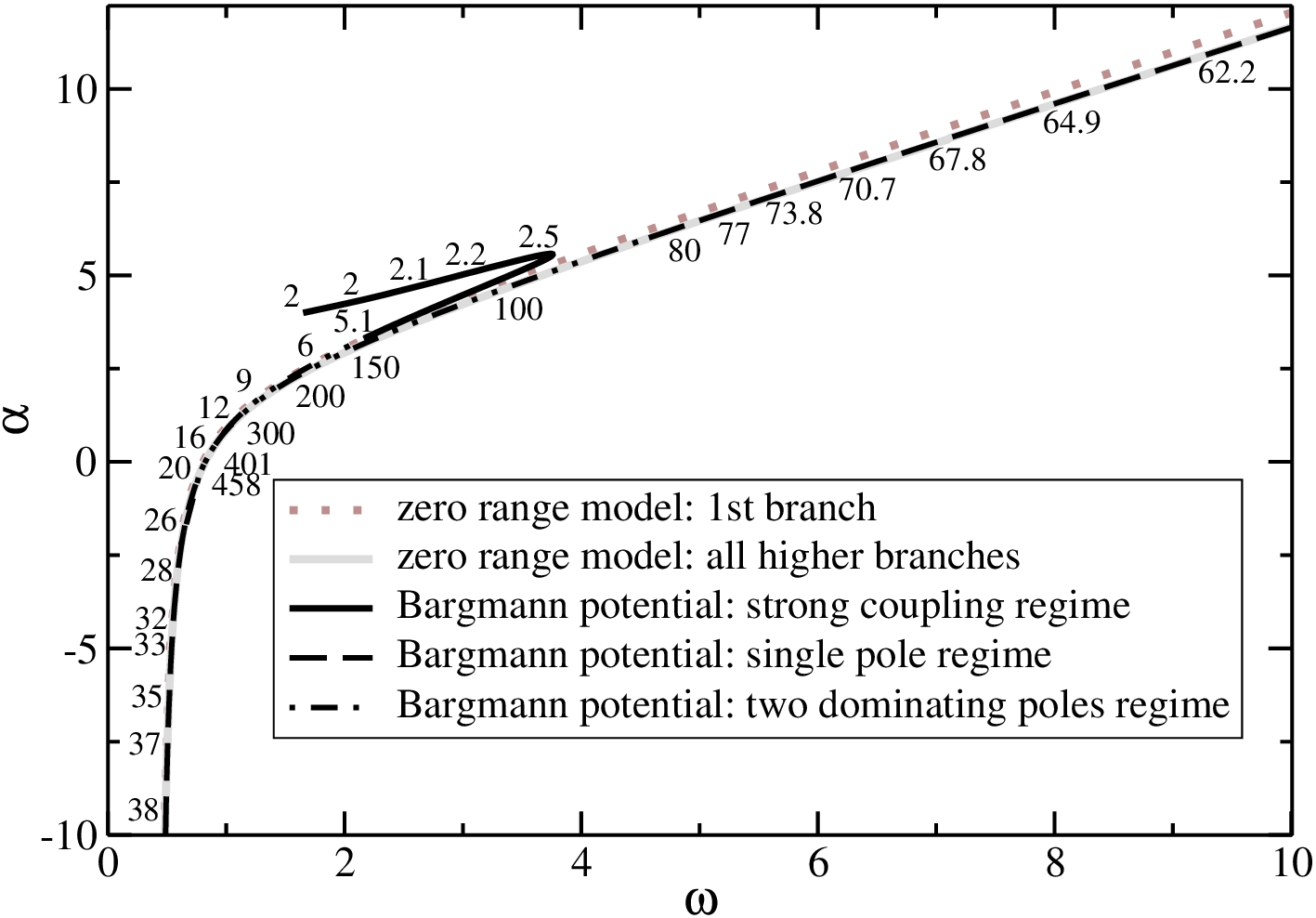}
  \caption{\protect\narrowtext
   The modified Phillips line for the Bargmann potential
   and the zero-range model. For the Bargmann potential the
   effective range is held fixed at 1 a.u., while the scattering
   length varies from 2 to 460 a.u. . The values of the two-body
   scattering length $a_2$ are shown along the curve.}
  \label{fig4}
\end{figure}

To better understand the complex correlation revealed
in Figure \ref{fig3}, we show in Figure \ref{fig4} the same correlation plot
using a single two-body potential (the Bargmann potential
with the effective range $r_0=1$ a.u.), but with large variations
of the two-body scattering length. We have also shown the three-body data
generated by solving the regularized Skornyakov/Ter-Martirosyan equations (STM)
\cite{STM,Kharchenko} corresponding to a zero-range interaction model.

We can identify two distinctive regimes: a universal regime
and a strong coupling regime.
In the strong coupling regime (characterized by relatively
small two-body scattering length) the shape of the modified
Phillips line depends on the potential (inset Fig. \ref{fig3} ) and,
typically, forms an elongated half-loop above the universal
curve. First, as the two-body scattering length
increases, the points on the Phillips line move right along the correlation plot reaching
a local maximum, and then turn left as they approach the universal regime.

In the universal regime, as the two-body scattering
length increases, the points along the Phillips line move
left along the universal curve. The shape of the universal
correlation plot stems from the interplay of poles of the
three-body $t$-matrix corresponding to the formation of near-threshold bound and virtual states.

We can distinguish three characteristic parts of the universal curve.
The linear part at the right end corresponds to large positive particle-dimer scattering length
and a very shallow three-body bound state.
The diverging part at the left end of the correlation plot corresponds to large negative
particle-dimer scattering length and indicates the formation
of a near-threshold virtual state. With increasing two-body scattering length
this virtual state turns into a bound state, and the corresponding point
on the modified Phillips line jumps to the far end of the positive linear part of the
universal curve. There is also a transitional nonlinear regime,
when the scaled particle-dimer scattering length $\alpha$ is small,
and both poles contribute to the shape of the universal curve.

The most transparent interpretation for the shape of the universal
part of the correlation plot can be obtained from a simple analysis of
the spectrum of the Faddeev operator
\[
\hat{K}(E)=\hat{G}_2(E) V (\hat{P}^+ + \hat{P}^-) \ ,
\]
where $\hat{G}_2(E)=(\hat{H}_0+V-E)^{-1}$ is the Green's function
for the three-body cluster Hamiltonian, $\hat{H}_0$ is
the Hamiltonian for three free particles,
 $V$ is the two-body pairwise potential and $\hat{P^{\pm}}$
are Jacobi coordinate transformation operators.
The equation for the component of the scattering wave function
\footnote{The three-body wave function can be recovered from the component as
$\Psi=(1+ P^+ + P^-)\Phi$.}
then reads
\[
[1+\hat{K}(E)]\Phi=-V (\hat{P}^+ + \hat{P}^-)\chi_0 \ ,
\]
where $\chi_0$ stands for the atom-dimer plane wave.
The component of the bound state wave function satisfies the homogeneous equation
\[
[1+\hat{K}(E)]\Phi=0 \ .
\]
The Faddeev operator $\hat{K}(E)$ for short-range potentials has a discrete spectrum
with eigenvalues $\lambda_n(E)$. For $E \le E_2$ the eigenvalues $\lambda_n(E)$ are real.

At the two-body threshold $E=E_2$ the Faddeev operator can be approximated
by a sum of projectors on the states corresponding to eigenvalues
$\lambda^+$ and $\lambda^-$ closest to $-1$, and
the particle-dimer scattering length can be expressed as
\begin{equation}
  a_3=\frac{a^+}{1+\lambda^+(E_2)} + \frac{a^-}{1+\lambda^-(E_2)} \ .
  \label{eq:twoPoles}
\end{equation}
There are two possible situations: 1) one of the eigenvalues -- $\lambda^+(E_2)$
or $\lambda^-(E_2)$ -- is very close to the critical value $\lambda=-1$
and the corresponding term gives the major contribution to the particle-dimer
scattering length; 2) both of the terms contribute substantially.
The first case is responsible for the ``linear'' and ``singular''
parts of the $\alpha(\omega)$ universal correlation plot.
The second case corresponds to the intermediate regime of small
scaled particle-dimer scattering length $\alpha$.

\begin{figure}
  \includegraphics[clip=true,width=0.5\textwidth]{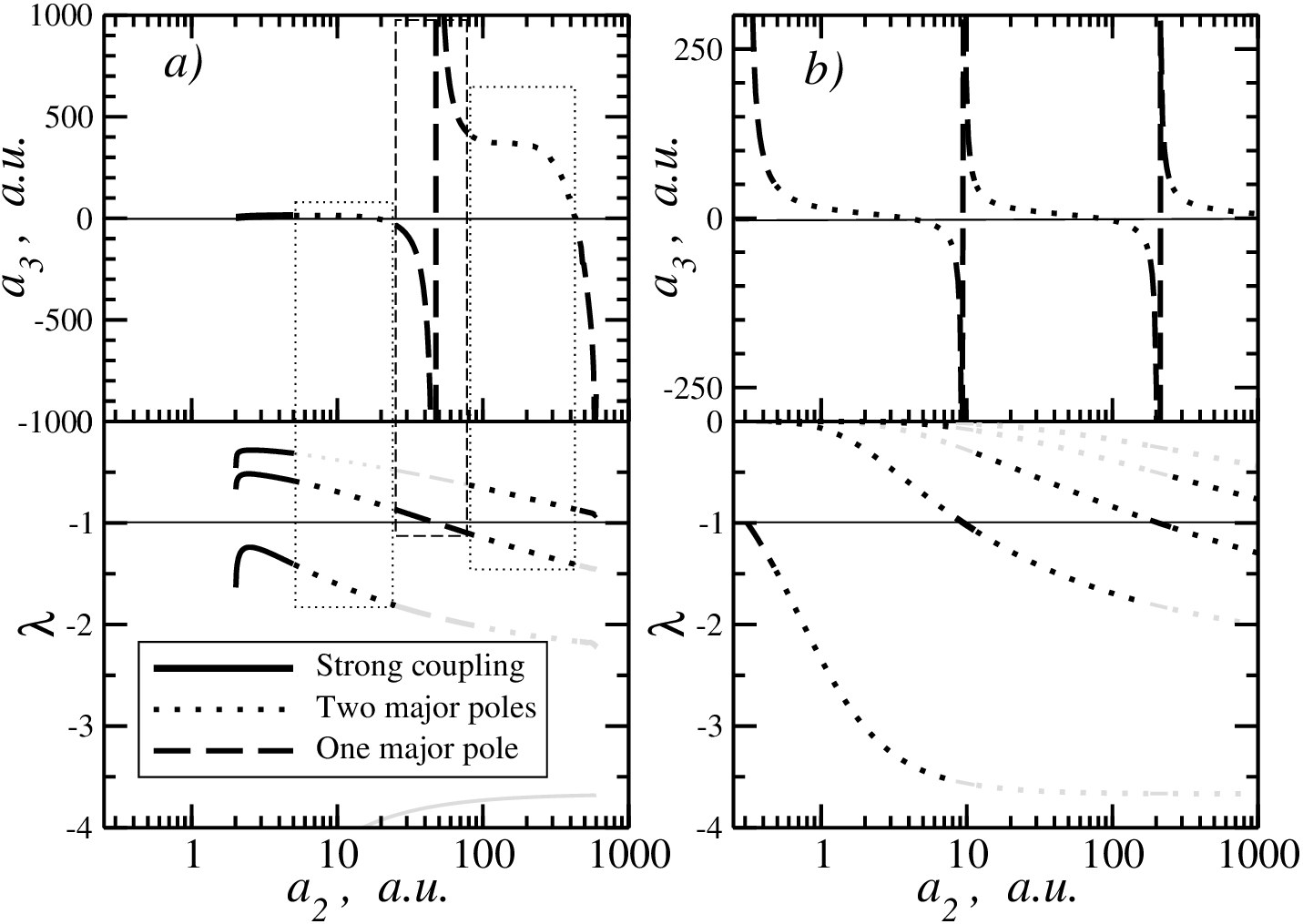}
  \caption{\protect\narrowtext
            Atom-dimer scattering length and the spectrum of the
            kernel of the Faddeev (a) or STM (b) equation at
            the two-body threshold as a function of the two-body
            scattering length. a) Bargmann potential with $r_0 = 1$ a.u.,
            b) zero-range interaction model.
     }
  \label{fig5}
\end{figure}

This interpretation is illustrated in Fig.~\ref{fig5}, where
we show the particle-dimer scattering length (top) and
the three-body operator spectrum (bottom) as a function of
the two-body scattering length. The poles in the three-body scattering
length correspond to the eigenvalue of the three-body operator
crossing the critical value $\lambda(E_2)=-1$, where a new three-body
bound state is formed.

The universal part of the modified Phillips line is reproduced by the simple formula
\begin{equation}
  \alpha=\frac{\alpha_1}{\frac{1}{\omega}-\frac{1}{\omega_0}} +\omega+\alpha_0
  \label{eq:parametrization}
\end{equation}
with $\alpha_1=5.5$, $\omega_0=0.419$ and $\alpha_0=4$. The first term here is responsible for the description of the large
negative scattering length regime, the other two terms fit the large positive scattering length regime.
Equation~\ref{eq:parametrization} can be used in practical calculations to estimate the trimer binding energy from the scattering length. A simpler empirical fit $\alpha=\omega+\frac{3}{2}(1-1/\omega^2)$ is suitable for positive three-body scattering length.

Careful analysis of the modified Phillips line reveals that in the universal, potential-independent regime the scattering length for the atom-dimer collision approaches zero when $E_3/E_2\approx 2.54$. The other special point on the plot corresponds to divergent negative scattering length,
which corresponds to the universal ratio $E_3/E_2\approx 6.7$.

The result demonstrates a novel form of universality in weakly bound systems. Simple in physical nature, the
modified Phillips line provides an important test for numerical and theoretical analysis of 3-body systems.
It affords the opportunity to check bound-state and scattering results for consistency and classifies
3-body systems according to distinct dynamical regimes.
It also provides an opportunity to check estimates of 3-body bound states and atom-dimer scattering lengths for internal consistency;
such as those obtained with regularized zero-range potential models \cite{Braaten}.
Correlation plots similar to the modified Phillips line for bosons can be constructed for three-body systems with nontrivial
spin-isospin structure and nonidentical particles. Here we shall only mention that the data shown in Figure \ref{fig1} is consistent
with the universal part of the modified Phillips line constructed on the base of the zero-range interaction model \cite{Kharchenko}.

All the calculations presented here have been performed using an original code for solving Faddeev equations
\cite{GridConst,Benchmark1,Benchmark2}. The code is currently capable of three-body bound state and scattering calculations below the
first excitation threshold. It provides users a simple interface for preparing configuration files and for constructing optimized grids automatically \cite{GridConst}.
The computational kernel solves the system of Faddeev equations for bound or scattering states. It uses a numerically effective computational scheme which -- in combination with an option of using multiple CPUs -- makes it possible to calculate scattering amplitudes quickly (from a few seconds to a few minutes, depending on the desired numerical accuracy and physical parameters of the system).

The code is available from the authors by request and will be available online in the near future.

\section{Acknowledgements}
Authors thank E. Kolganova for stimulating discussions and independent testing of our three-body code.
This work is supported by NSF grant PHY-0903956.

\end{document}